\documentclass{elsart}
\usepackage{graphicx}

\begin{document}
 
\begin{frontmatter}   
 
\title{Loss of  superfluidity  in a Bose-Einstein
condensate via forced resonant oscillations}
 
\author{Sadhan K. Adhikari}\footnote{{\it E-mail address:}
adhikari@ift.unesp.br}

\address{Instituto de F\'{\i}sica Te\'orica, Universidade Estadual
Paulista, \\ 01.405-900 S\~ao Paulo, S\~ao Paulo, Brazil}

\date{\today}
 \maketitle

\begin{abstract}

We predict the loss of superfluidity in a Bose-Einstein condensate in an
axially symmetric harmonic trap alone during resonant collective
oscillations via a classical dynamical transition. The forced resonant
oscillation can be initiated by (a) a periodic modulation of the atomic
scattering length with a frequency that equals twice the radial trapping
frequency or multiples thereof, or by (b) a periodic modulation of the
radial trapping potential with a frequency that equals the radial trapping
frequency or multiples thereof.  Suggestion for future experiment is made.

\end{abstract}


\end{frontmatter}

The first experimental observation  of quantum phase
effects on a macroscopic scale such as interference of matter
waves \cite{kett1} was made in  
a Bose-Einstein condensate 
(BEC) in an  axially-symmetric harmonic 
trap. Later on more controlled studies of interference of matter waves
were performed with BEC loaded on a harmonic plus an one-dimensional
optical lattice trap \cite{1,2,cata,catax,cata3,xxx}. More recently, a
three-dimensional optical
lattice trap has been employed by Greiner {\it et al.}
\cite{greiner,stoof,adhi7} in the study of
the formation of an interference pattern.   The formation of the
interference pattern is a consequence of phase coherence in the BEC across
the optical lattice sites.

In the experiments with optical lattice,  phase coherence in the BEC is
generated by quantum tunneling of atoms
from one optical lattice site to another leading to a communication among
various sites. As the strength of the optical lattice  barrier is  larger
than the typical energy of the system by about two orders of magnitude,
classical
movement of uncondensed cold atoms  is prohibited across the
optical lattice barriers. It is the atoms of the BEC which experience this
miraculous flow through  the high barriers, which is a manifestation of 
superfluidity. 
More surprising is the observation that 
all superfluids are necessarily condensates, but all condensates are not
superfluids. There have been both theoretical and experimental  studies of 
breakdown of superfluidity in BEC via quantum
\cite{greiner,greiner1} and classical \cite{sm,cata2,adhi2,adhi3} 
transitions.

As superfluidity is necessarily a quantum phenomenon the experimental
consideration of Greiner {\it et al.} \cite{greiner,stoof,greiner1} on the
loss of superfluidity in a
BEC trapped on a three-dimensional optical lattice potential via a quantum
phase transition is worth mentioning.   The long-range
phase coherence in the condensate along the entire optical lattice is a
sign of  communication among various sites which is necessary for
developing superfluidity in the condensate. Equal phase at all points or a
slowly (and orderly) varying phase are the ideal examples
of coherent phase. On the other hand, a rapidly (and randomly)  varying
phase in space is generally incoherent. The phase on an optical lattice
site and the number of atoms in that
site play the roles of conjugate variables obeying the Heisenberg
uncertainty principle of quantum mechanics \cite{stoof}.  In the
superfluid state the coherent phase is considered to be known and
consequently the number of atoms on each site is unknown thus allowing
a free movement of atoms from one site to another \cite{greiner}. 
The loss of phase coherence is associated with a
Mott insulator state where  the phase is entirely arbitrary across the
optical lattice sites  and the number of
atoms at each site is fixed and consequently, their free passage from one
site to
another is stopped.

Greiner {\it et al.}
\cite{greiner,greiner1} demonstrated  that, as the strength of the optical
lattice
traps
is increased, the quantum tunneling of atoms from one optical site to
another as well as the communication among different optical lattice sites 
are 
stopped resulting in a superfluid to Mott insulator quantum phase
transition in the BEC \cite{greiner}. As the
 strength of  the optical lattice  traps  in the  Mott insulator state is
reduced the superfluidity is restored in a short time 
\cite{greiner,greiner1}. This reversible quantum
phase transition may occur at absolute zero (0 K) and is driven by
Heisenberg's uncertainty principle \cite{greiner} and not by thermal
fluctuations  
involving  energy
as in a
classical phase transition. As the temperature approaches absolute zero
all thermal fluctuations die out and at 0 K classical phase transitions
are necessarily
excluded.

Following a suggestion by Smerzi {\it et al.} \cite{sm}, Cataliotti {\it
et al.} \cite{cata2} demonstrated in a novel experiment the
loss of superfluidity in a BEC trapped in an one-dimensional
optical-lattice and harmonic potentials when the center of the harmonic
potential is suddenly displaced along the optical lattice through a
distance larger than a critical value.  Then a modulational instability of
classical nature
takes place in the BEC. Consequently, it cannot reorganize itself quickly
enough via phase coherent states
and the superfluidity of the BEC is lost.  
Later on, it has been suggested  that superfluidity could be
lost in a resonant collective oscillation of a BEC on an  one-dimensional
optical lattice potential arising from a periodic modulation of the
scattering length \cite{adhi2} or a  periodic modulation of the   radial
trapping potential \cite{adhi3}.

All the above studies on the destruction of superfluidity in a BEC were
performed with an optical lattice trap and the loss of superfluidity
in both classical and quantum cases has been traced to the
fixing of a specific atom(s) at a definite lattice site with no mobility
 \cite{greiner,stoof,sm}. An
interesting question is if, in case of the loss of superfluidity via the 
classical transitions above \cite{adhi2,adhi3},  the forced tunneling of
the atoms through
the optical lattice barriers  plays a fundamental role. 
We find that the answer to this question is negative and 
demonstrate that the
superfluidity  can also be destroyed in a  classical
resonant oscillation of a BEC in a harmonic trap alone without any
accompanying optical lattice trap. In view of comments in the literature
\cite{cata,greiner,stoof} this is surprising.
Effectively,  the
optical lattice  is found to have  no effect on the loss of mobility of
the
atoms associated with  quantum fluctuations due to Heisenberg's
uncertainty principle. 

In the present study the forced classical
resonant oscillation of the BEC is initiated by a periodic modulation of
the
scattering length or the radial trapping potential. A periodic modulation
of
the
scattering length \cite{adhi4}  or the radial trapping potential
\cite{osc,osc1} is
known to
lead to
collective resonant oscillation in an axially-symmetric  BEC.
In  previous studies we
demonstrated that such forced
oscillations in the presence of a joint harmonic and optical traps lead
to
a breakdown of superfluidity \cite{adhi2,adhi3}. A similar breakdown of
superfluidity of
a BEC in a harmonic trap alone reveals that  the optical
lattice trap does not play a decisive role in the loss of superfluidity
via collective resonant oscillation.  The optical lattice trap is,
however, fundamental in the formation of the interference pattern upon
release from the traps, which plays a decisive role in the 
detection of
superfluidity in these studies.
The strong optical lattice potential essentially cuts
the BEC into phase coherent pieces upon release from the joint traps which
is fundamental  in the formation of the interference pattern.

Another way of detection of phase coherence and superfluidity of a
BEC is by cutting it into phase coherent pieces  using a laser as in the
experiment by 
Andrews {\it et al.} \cite{kett1}. Upon free expansion these
pieces form
an  interfence pattern. 
If the BEC is cut into two pieces the interference
pattern consists of a large number of dark (absence of matter) and bright
(matter) patches \cite{kett1}. However, we show that if the initial BEC is
cut into
many equal
coherent pieces as in the experiment with optical lattice,
only three prominent bright spots can be generated. In the present
numerical simulation with an axially symmetric harmonic trap, the initial
wave function $\psi$  of the BEC is cut into many equal pieces by setting 
$\psi(\rho, z= \pm j d)=0$ with $j$ = 0, 1, 2,
3,...etc. Here
$d$ is the spacing in the axial $z$ direction and $\rho$ corresponds to  
the
radial direction. Subsequently, for a sufficiently small $d$ ($\sim 0.1 
- 1 \mu$m), 
upon release
from the trap such a coherent
fragmented BEC will form a interference pattern of three bright 
spots.

We consider here two ways of initiating the collective resonant
oscillation of the BEC.
First, it   is initiated near a
Feshbach resonance \cite{mit} by a
periodic modulation   of the repulsive atomic scattering length $a$ $ 
(>0)$
 via $a \to a +\bar A\sin(\Omega t)$ where
$t$ is time, $\bar A$ an amplitude, 
and $\Omega$ the frequency of modulation.  
Such modulation of the  scattering length  can be realized
experimentally near a Feshbach
resonance by manipulating an external background magnetic field. 
When  $\Omega= 2\omega$ or  multiples thereof, 
resonant collective oscillation can be generated  in the BEC,  where
$\omega$ is the 
radial trapping frequency \cite{adhi4}. This resonant oscillation 
destroys the superfluidity of the BEC provided that the condensate is
allowed to experience this oscillation for a certain interval of
time called hold time. 

Next we generate the resonant collective oscillation by a periodic
modulation of the radial trapping potential $V_\rho$ in the axially
symmetric
BEC via $V_\rho \to V_\rho[1+\bar B\sin (\Omega t)]$ with $\bar B$ the
amplitude
of modulation.  
When  $\Omega= \omega$ or  multiples thereof,
resonant collective oscillation can be generated  in the BEC
\cite{osc}. This phenomenon has also been explored experimentally
\cite{3}.
We find that this resonant oscillation also destroys
superfluidity of the BEC after some hold time. 

The transition we consider is classical in nature and can be treated
\cite{sm}
by the  mean-field nonlinear Gross-Pitaevskii (GP) equation \cite{8}.  
The time-dependent axially symmetric BEC wave
function $\Psi({\bf r};t)$ at position ${\bf r}$ and time $t $
is described by the following   GP equation
\begin{eqnarray}\label{a} \biggr[- i\hbar\frac{\partial
}{\partial t}
-\frac{\hbar^2\nabla^2   }{2m}
+ V({\bf r})
+ gN|\Psi({\bf
r};t)|^2
 \biggr]\Psi({\bf r};t)=0,\end{eqnarray}
where $m$ is the mass and  $N$ the number of atoms in the condensate,
 $g=4\pi \hbar^2 a/m $ the strength of interatomic interaction, with
$a$ the atomic scattering length, and       $  V({\bf
r}) =\frac{1}{2}m \omega ^2(\rho ^2+\nu^2 z^2) $ the trapping potential 
where
 $\omega$ is the angular frequency of the harmonic trap
in the radial direction $\rho$, $\nu \omega$ that in  the
axial direction $z$, with $\nu$ the aspect ratio. 
The normalization condition  is $ \int d{\bf r} |\Psi({\bf r};t)|^2 = 1. $

In the axially-symmetric configuration, the wave function
can be written as  $\Psi({\bf r}, t)= \psi(\rho ,z,t)$.
Now  transforming to dimensionless variables $\hat \rho =\sqrt 2 \rho /l$,
$\hat z=\sqrt 2 z/l$, $\tau=t
\omega, $ $l\equiv \sqrt {\hbar/(m\omega)}$,
and ${ \varphi(\hat \rho,\hat z;\tau)} \equiv   \hat \rho
\sqrt{{l^3}/{\sqrt
8}}\psi(\rho ,z;t),$  Eq.  (\ref{a}) becomes 
\begin{eqnarray}\label{d1} &\biggr[&-i\frac{\partial }{\partial \tau}
-\frac{\partial^2}{\partial \hat \rho^2}+\frac{1}{\hat
\rho}\frac{\partial}{\partial \hat \rho} -\frac{\partial^2}{\partial \hat
z^2} +\frac{1}{4}\left(\hat \rho^2+\nu^2 \hat z^2\right) \nonumber \\
&+&\frac{V_{\mbox{opt}}}{\hbar \omega} -{1\over \hat \rho^2} + 8\sqrt 2
\pi n\left|\frac {\varphi({\hat \rho,\hat z};\tau)}{\hat \rho}\right|^2
 \biggr]\varphi({ \hat \rho,\hat z};\tau)=0,
\end{eqnarray}
where nonlinearity
$ n =   N a /l$. In terms of the
one-dimensional probability
 $P(z,t) \equiv 2\pi$ $\int_0 ^\infty
d\hat \rho |\varphi(\hat \rho,\hat z,\tau)|^2/\hat \rho $, the
normalization
of 
the
wave
function
is given by $\int_{-\infty}^\infty d\hat z P(z,t) = 1.$
The probability
$P(z,t)$ is  useful in the study of the present problem under the
action of the optical lattice, specially in the
investigation of the formation and
evolution of the interference pattern after the removal of the
trapping potentials.

We use the parameters of 
the  experiment of Cataliotti {\it et al.} \cite{cata}
with repulsive $^{87}$Rb atoms where  
the radial trap frequency was  $ \omega =
2\pi \times 92$ Hz. 
For the mass $m=1.441\times 10^{-25}$ kg of $^{87}$Rb the
harmonic
oscillator length $l=\sqrt {\hbar/(m\omega)} = 1.126$ $\mu$m and 
and the 
dimensionless time unit  $\omega ^{-1} =
1/(2\pi\times 92)$ s $ =1.73$ ms. 
We solve  Eq.  (\ref{d1}) numerically  using a   
split-step time-iteration
method
with  the Crank-Nicholson discretization scheme described recently
\cite{11}.  
The time iteration is started with the known harmonic oscillator solution
of  Eq.  (\ref{d1}) with
 $n=0$: $\varphi(\hat \rho ,\hat z) = [\nu
/(8\pi^3)  ]^{1/4}$
$ \hat \rho e^{-(\hat \rho ^2+ \hat z ^2)/4}$ 
\cite{11}. 
First we calculate  the ground-state wave
function of the system for $n=5$ and $\nu =0.5$ which we use in numerical 
simulation. 
 To investigate the superfluidity of the BEC
we set  $\varphi(\hat \rho ,\hat z)=0$ for $\hat z=\pm 0.3j$, $j=0, 1, 2,
3,...$. This cuts the initial BEC in slices of width 0.3 in the axial
direction. The free expansion of  a phase coherent initial BEC sliced in
this fashion will lead to  a prominent interference pattern 
revealing the superfluidity.   Essentially,
similar slices result in a BEC trapped in an
one-dimensional optical lattice as in the experiment by Cataliotti {\it 
et al} \cite{cata}.   
The atom
cloud released from one piece
expand, and overlap and interfere with atom clouds from neighboring
piece to form the robust interference pattern due to phase coherence.
The pattern consists of a central peak 
and two symmetrically spaced peaks
moving in opposite directions \cite{cata,catax,xxx}.
Although we employ the dimensionless space units $\hat \rho$ and $\hat z$
and
time unit
$\tau$ in numerical calculation, the results are reported in actual units:
$r$ $\mu$m, $\rho$  $\mu$m,  $z$ $\mu$m  and $t$ ms. In the conversion we
used the
parameters of the experiment of Cataliotti {\it et al.} \cite{cata}, e.g.,
$\rho= 0.8\hat \rho$ $\mu$m, $z=0.8\hat z$ $\mu$m, and $t=1.73\tau$ ms.
 
\begin{figure}[!ht]
 
\begin{center}

\includegraphics[width=.75\linewidth]{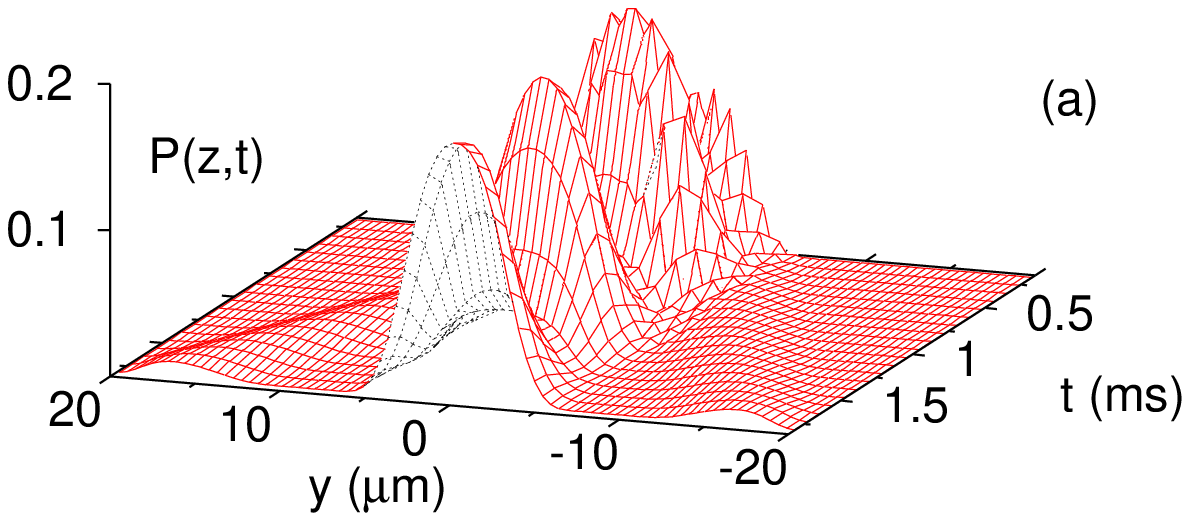}
\includegraphics[width=.75\linewidth]{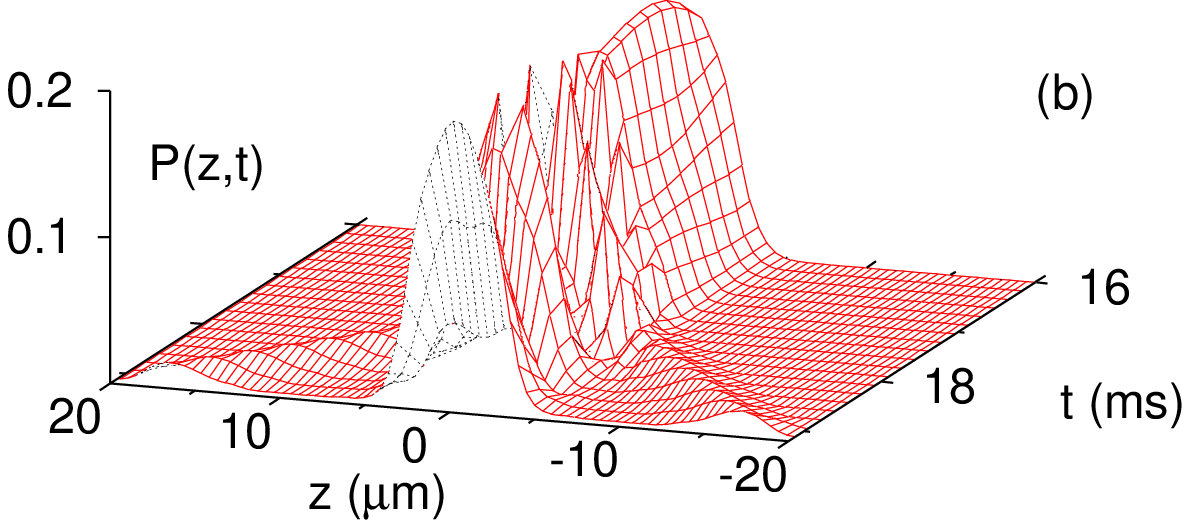}
\includegraphics[width=.75\linewidth]{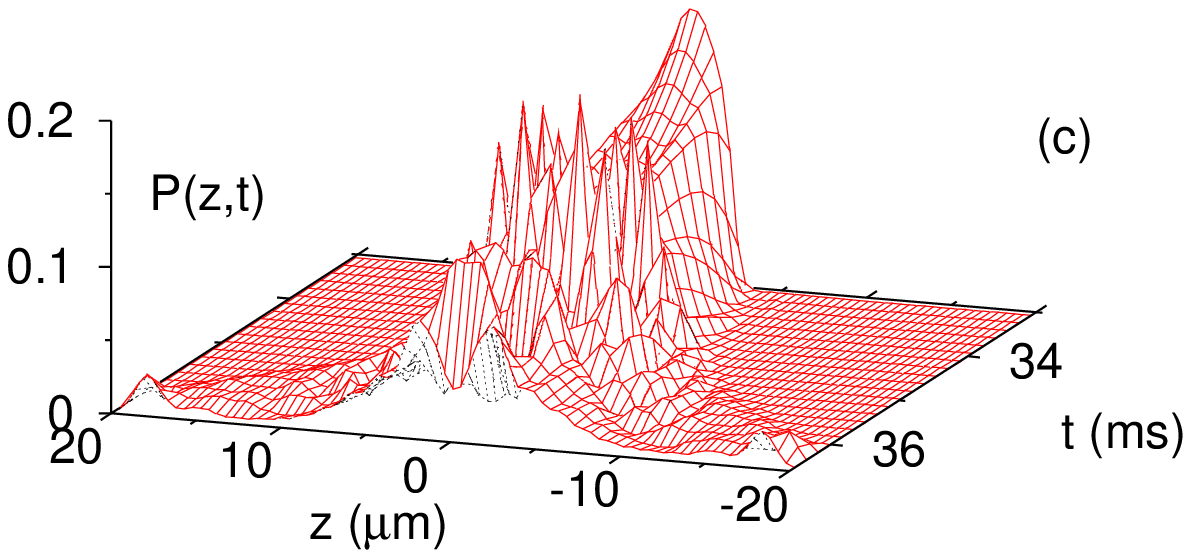}
\includegraphics[width=.75\linewidth]{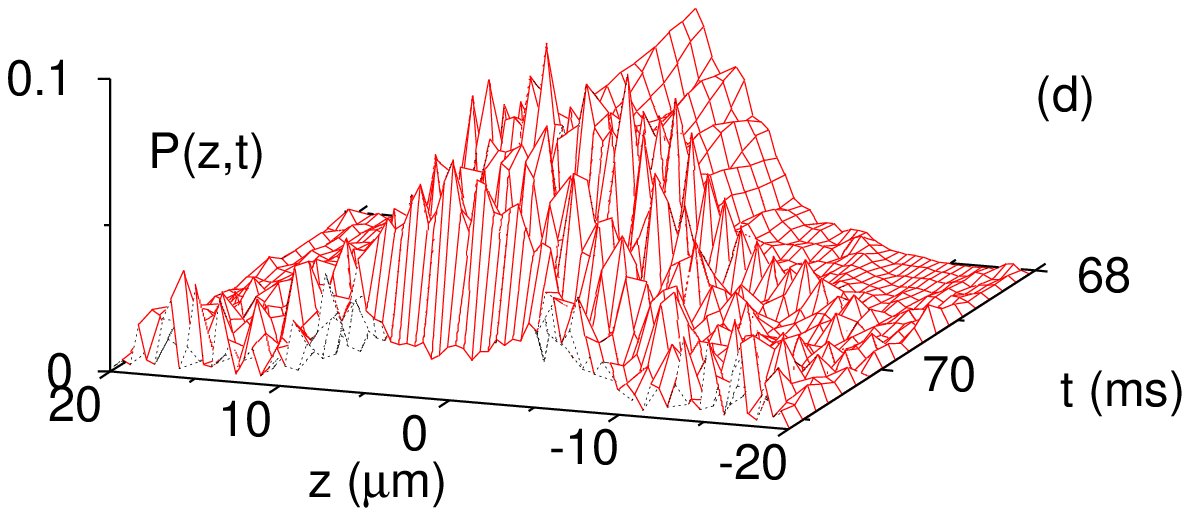}
\end{center}
 
\caption{One-dimensional probability  $P(z,t)$
vs. $z$ and $t$ for the BEC  
under the action of modulation (\ref{rep}) of the nonlinearity with
$n=5$, $\Omega =
2\omega$,
and $A=3$ 
and upon the removal of the
traps after hold times (a) 0, (b) 17 ms,  (c) 35 ms, and
(d) 69 ms.  
} \end{figure}

First we study the 
destruction of superfluidity in the condensate  after the application of a
periodic modulation of 
the scattering length resulting in a similar modulation of nonlinearity
$n$ in
Eq. (\ref{d1}) via 
\begin{equation}\label{rep}
n  \to n +A\sin (\Omega \tau),
\end{equation}where $A$ is an amplitude.  
In the present study
we employ  amplitude $A=3$. 
For present $n=5$, this will restrict the modulated nonlinearity to
have always positive values corresponding to atomic repulsion. Negative
values of nonlinearity corresponding
to atomic attraction may lead to collapse and instability \cite{8,adx}
and will not be considered here. The
BEC
has been found to execute
resonant collective oscillation when the modulation frequency $\Omega$
is an even multiple of the radial trap frequency $\omega$ \cite{adhi4}. In
the presence
of an one-dimensional optical lattice trap such resonant nonlinear
oscillation
destroys superfluidity \cite{adhi2}. Here we investigate
if the
forced
quantum tunneling of the atoms across the optical lattice trap during
classical oscillation of the BEC is responsible for the loss of
superfluidity.

\begin{figure}
 
\begin{center}

\includegraphics[width=\linewidth]{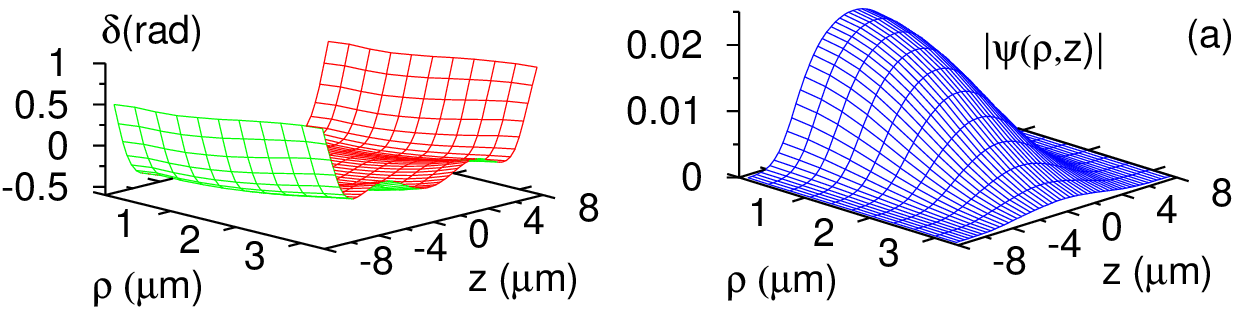}
\includegraphics[width=\linewidth]{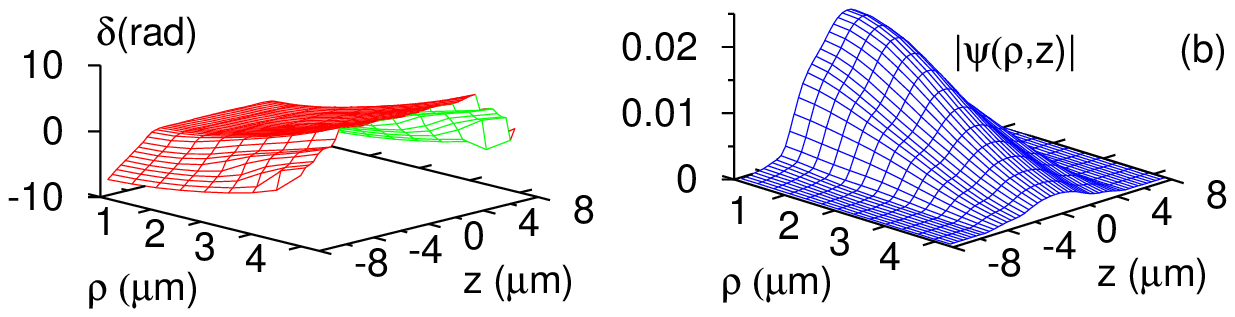}
\includegraphics[width=\linewidth]{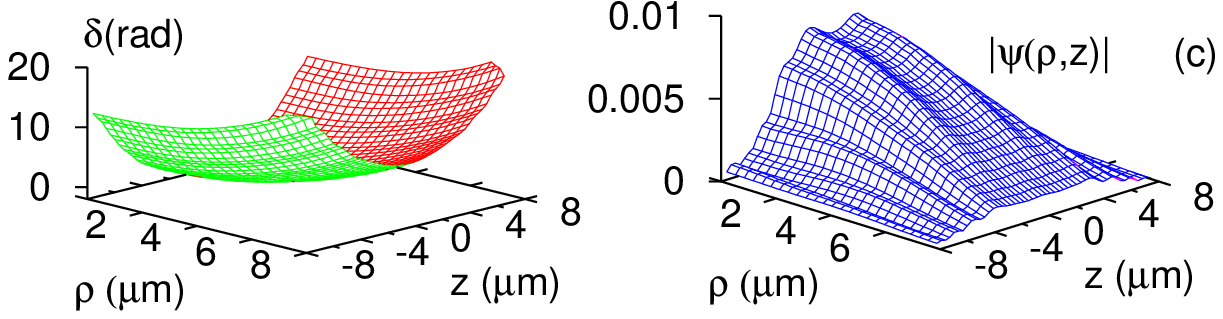}
\includegraphics[width=\linewidth]{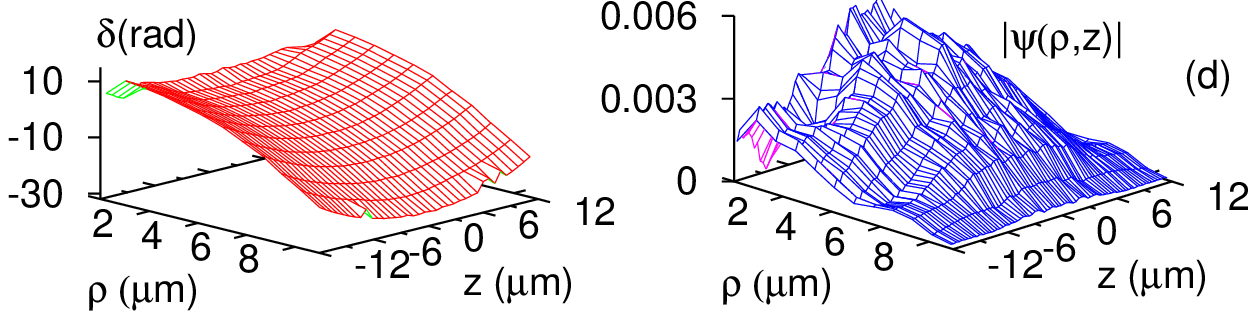}
\end{center}
 
\caption{Phase $\delta$ (left) and modulus $|\psi(\rho,z)|$ (right) of the 
wave function 
vs. $\rho$ and $z$ for the BEC  
under the action of modulation (\ref{rep}) of the nonlinearity with
$\Omega =2\omega$,
 $n=5$ and   $A=3$ after hold times (a) 0, (b) 17 ms, (c) 35 ms, and
(d) 69 ms.
} \end{figure}

We explicitly study the destruction of superfluidity in the condensate
upon the application of  modulation (\ref{rep}) of the scattering length  
leading to a resonant
oscillation at $\Omega=2\omega$. The loss of superfluidity only takes
place if the BEC is
allowed to experience the resonant oscillation  for a
certain hold time.  
For numerical simulation we allow the BEC to evolve on  a lattice with 
$\rho \le 20$ $\mu$m and 20 $\mu$m $\ge z\ge$ $-20$ $\mu$m after the
modulation (\ref{rep}) is applied 
and study
the system after different   hold times.
The one-dimensional
probability  $P(z,t)$ is  plotted in Figs. 1 (a), (b), (c) and (d)
for hold times 0, 17 ms,  35 ms and  69 ms,  respectively. For 
hold
time  17 ms,  prominent interference
pattern is
formed upon free expansion.
In Fig.  1 (a)  
three separate piece in the  interference pattern  
corresponding  to three distinct trails can be identified. 
The
interference 
pattern is slowly destroyed at  increased  hold times as  we can see in
Figs.  1 (b),  (c) and (d). 
As the hold time  increases the maxima
of the interference pattern
mix up upon free expansion  and finally for the  hold time of 
69 ms  the interference pattern is
completely destroyed as we find in Fig.  1 (d).
As the BEC is allowed to evolve for a substantial  interval of time after
the application of the periodic modulation in the scattering length, a
dynamical instability of classical nature sets in which destroys the
superfluidity \cite{sm,cata2}.

Next we investigate how the phase $\delta$ over the condensate behaves as
the hold time is increased and what happens to the modulus
$|\psi(\rho,z)|$ of the wave function. For this purpose we plot the phase
and the modulus of the wave function of the BEC's of Figs. 1 (a), (b),
(c), and
(d), respectively, 
in Figs. 2 (a), (b), (c), and
(d) after  hold times 0, 17 ms, 35 ms, and 69 ms.  
For hold time 0 the wave function is smooth and the phase is slowly
varying over the full condensate corresponding to perfect phase coherence. 
But as the hold time is increased the phase varies rapidly over the
condensate. The total variation of phase over the BEC for hold times 0, 17
ms, 35 ms, and 69 ms are 1.5 rad, 15 rad, 20 rad, and 45 rad,
respectively. For larger hold times  the modulus
$|\psi(\rho,z)|$ of the wave function also develops a nonsmooth irregular
behavior via the  dynamical instability.
The rapid
variation of $\delta$ and the  nonsmooth irregular wave function 
over the BEC for larger hold times are  responsible for 
the  destruction of superfluidity.  
 Hence  the superfluidity in the
BEC could be
destroyed in the absence of an optical trap due a classical collective
resonant oscillation  in the condensate initiated by a periodic modulation
of the
scattering length with a frequency $\Omega = 2\omega$. Such resonances
appear for  $\Omega = 2\omega j$ \cite{adhi4}, where $j = 1, 2, 3,...$ and
we have
verified that similar breakdown of superfluidity also occurs for  $\Omega
= 4\omega$. The resonance becomes  more vigorous  as $A$ is increased and
so is the destruction of superfluidity.

\begin{figure}[!ht]
 
\begin{center}

\includegraphics[width=.9\linewidth]{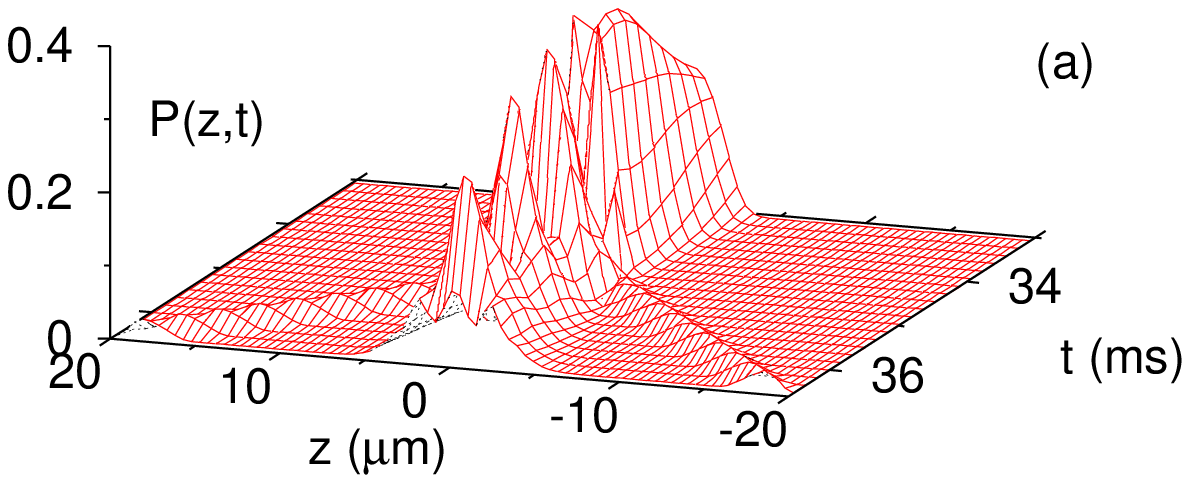}
\includegraphics[width=.9\linewidth]{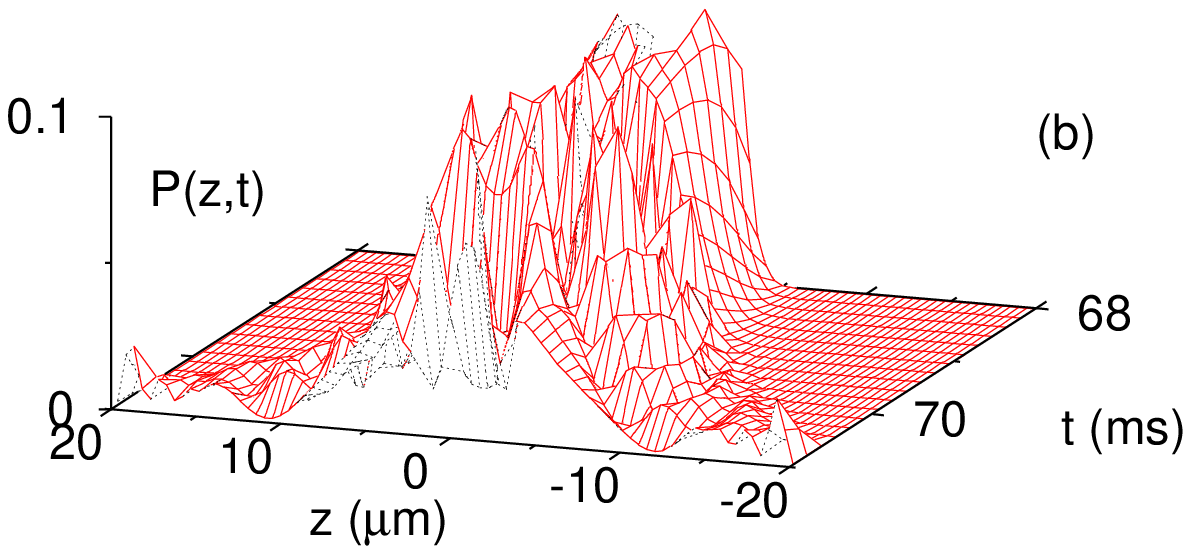}
\includegraphics[width=.9\linewidth]{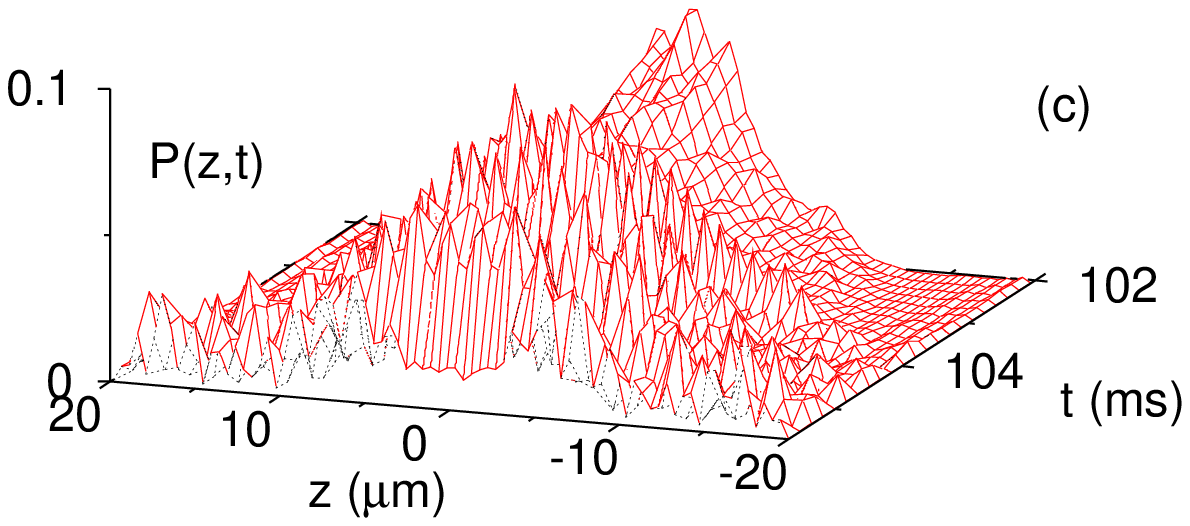}
\end{center}
 
\caption{One-dimensional probability  $P(z,t)$
vs. $z$ and $t$ for the BEC  
under the action of modulation (\ref{mod}) of the radial trapping
potential with $n=5$, $\Omega =
\omega$,
and $B=0.5$ 
and upon the removal of the
traps after hold times (a)    35 ms, 
(b) 69 ms, and (c) 104 ms.  
} \end{figure}

\begin{figure}[!ht]
 
\begin{center}

\includegraphics[width=.9\linewidth]{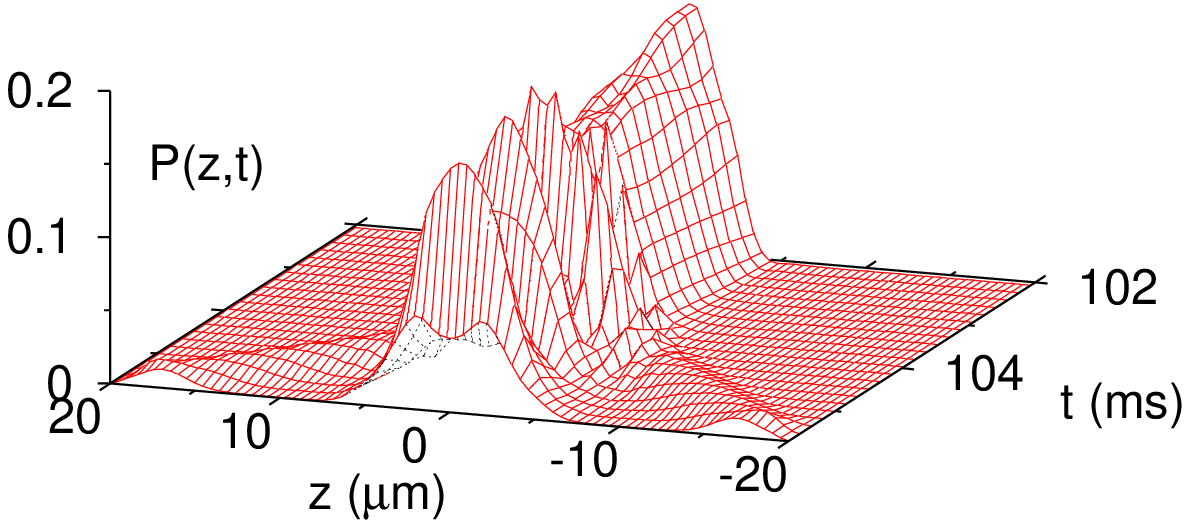}
\end{center}
 
\caption{One-dimensional probability  $P(z,t)$
vs. $z$ and $t$ for the BEC  
under the action of modulation (\ref{mod}) of the radial trapping
potential with $n=5$, $\Omega =
1.2\omega$,
and $B=0.5$ 
and upon the removal of the
traps after hold time 104 ms.  
} \end{figure}

However, resonant oscillation arising from modulation (\ref{rep}) is not
the only classical dynamical mechanism for the destruction of
superfluidity. It can also be destroyed  via a periodic
modulation of the
radial trapping potential \cite{osc}. 
To illustrate this we consider the following modulation of the radial
trapping potential in (\ref{d1}) via
\begin{equation}\label{mod}
\hat \rho^2 /4  \to (\hat \rho^2 /4 )[1+B\sin (\Omega \tau)],   
\end{equation}
while the axial trapping potential is left unchanged. In this case
parametric resonance appear when the modulation frequency $\Omega$ is a
multiple of the trap frequency $\omega$ \cite{osc}. These resonances
appear even for
a small  value of the amplitude $B$. However, they appear for a band of
frequency around the multiples of  $\omega$ for a larger value of $B$. 

We find that the superfluidity is destroyed if the modulation (\ref{mod})
of the  radial
trapping potential is continued for a certain hold time with $\Omega$
fixed at a resonant frequency. 
To illustrate the loss of superfluidity in this case we plot the
one-dimensional probability distribution $P(z,t)$ vs. $z$ and $t$ in
Figs. 3 (a), (b), and (c) for hold times 35 ms, 69 ms, and 104 ms,
respectively, for $\Omega =\omega$ and $B=0.5$.  In Fig. 3 (a) we find
that the superfluidity is maintained for a hold time of 35 ms and three
prominent peaks are formed in this case. However, as the hold time is
increased the clean interference pattern is gradually destroyed as can be
seen from Figs. 3 (b) and (c). The same phenomenon is found to occur 
at a larger 
resonant frequency of modulation, where  $\Omega$    is a larger multiple 
of  $\omega$. The destruction of superfluidity is favored for a larger
amplitude of modulation $B$.

If the frequencies  $\Omega$ are off their resonance values
and the amplitudes  $A$ and $B$ in Eqs. (\ref{rep}) and (\ref{mod}),
respectively,  have moderate values, there
is no loss of superfluidity after large hold times for  modulations
of scattering length or radial trapping potential. To illustrate
this for modulation (\ref{mod}) with $B=0.5$,  $\Omega =1.2\omega$ and a
hold time of 104 ms we plot the density $P(z,t) $ vs. $z$ and $t$ in
Fig. 4, where the clean interference pattern reappears 
with a small change in the modulation frequency $\Omega $ from the
resonance value $\omega$ to a nonresonant value $1.2\omega$.
The maintenance of superfluidity in Fig. 4 off the resonance should be
contrasted
with its loss in Fig. 3 (c) at resonance for the same hold time.

In conclusion, 
we studied  the destruction of superfluidity 
in a cigar-shaped trapped BEC   upon the application of 
a periodic modulation of the scattering length or of the radial trapping
potential 
leading to a resonant
collective excitation.   In the absence of 
modulation,  the
formation of the interference pattern  upon the removal of the
trap clearly demonstrates the
phase coherence. 
At the resonance frequencies a dynamical instability   
in the BEC leads to  the 
 destruction of superfluidity    
provided
that 
the BEC is kept in the modulated trap for a certain hold time.
Consequently,
after release from the trap no interference pattern is
formed. The superfluidity  in the BEC  is not destroyed  when the
frequency of
modulation  is changed to a nonresonant value. 
It is possible to study this novel 
way of the destruction of superfluidity
experimentally and a comparison of those results  with
mean-field models will 
enhance our understanding  of matter wave BEC.

\ack   

The work was supported in part by the CNPq and FAPESP
of Brazil.


\begin{thebibliography}{10} 




 
 
 
\bibitem{kett1}M. R. Andrews, C. G. Townsend, H. J. Miesner, D. S. Durfee,
D. M.  Kurn, W.
Ketterle,  Science { 275} (1997)  637.
   
 
 
 
 
\bibitem{1}  B. P. Anderson,  M. A. Kasevich, Science { 282} (1998) 1686.
 
 
\bibitem{2}C. Orzel, A. K.
Tuchman, M. L. Fenselau, M. Yasuda,  M. A. Kasevich,
Science { 291} (2001) 2386.

\bibitem{cata} F. S. Cataliotti, S. Burger, C. Fort, P. Maddaloni,
F. Minardi, A. Trombettoni, A. Smerzi,  M. Inguscio,
Science
{ 293} (2001) 843.
 
\bibitem{catax}P. Pedri, L. Pitaevskii, S. Stringari, 
   C. Fort, S. Burger, F. S. Cataliotti, P. Maddaloni, F. Minardi,
 M. Inguscio, Phys. Rev. Lett. 87 (2001) 220401.  

\bibitem{cata3}
 S. Burger, F. S. Cataliotti, C. Fort, F. Minardi, 
M. Inguscio,    M. L. Chiofalo, M. P. Tosi, Phys. Rev. Lett.
86 (2001) 4447;
 

\bibitem{xxx}J. H. M\"uller, O. Morsch, M. Cristiani, D. Ciampini,
E. Arimondo,  e-print cond-mat/0211079;

 O. Morsch, M. Cristiani, J. H. M\"uller,  D. Ciampini,  E. Arimondo,
Phys. Rev. A 66 (2002) 021601.
 
 
\bibitem{greiner}M. Greiner, O. Mandel, T. Esslinger, T. W. H\"ansch,
 I. Bloch, Nature (London) { 415} (2002) 39.
 

\bibitem{stoof}H. T. C.  Stoof,  {   Nature (London)} {
415} (2002) 25.

\bibitem{adhi7}S. K. Adhikari,  P. Muruganandam, Phys. Lett. A 310
(2003) 229.

\bibitem{greiner1}M. Greiner, O. Mandel, T. W. H\"ansch,
 I. Bloch, Nature (London) { 419} (2002) 51.


 
\bibitem{sm}  A. Smerzi, A. Trombettoni, P. G. Kevrekidis,  
A. R. Bishop, Phys. Rev. Lett.  { 89} (2002)  170402.

 
\bibitem{cata2} F. S. Cataliotti, L. Fallani, F. Ferlaino, C. Fort,
P. Maddaloni, M. Inguscio, A. Smerzi, A. Trombettoni, P. G. Kevrekidis,
A. R. Bishop, e-print cond-mat/0207139.



\bibitem{adhi2}S. K.   Adhikari, Phys. Lett. A 308 (2003) 302.

 
\bibitem{adhi3}S. K.   Adhikari,  e-print cond-mat/0211026.






\bibitem{adhi4}S. K.   Adhikari,  J. Phys. B 36 (2003) 1109;

 F. K. Abdullaev, J. C. Bronski, R. M. Galimzyanov, e-print
cond-mat/0205464. 

\bibitem{osc}J. J. Garc\'ia-Ripoll, V. M. P\'erez-Garc\'ia,
P. Torres, {
Phys. Rev. Lett.} { 83} (1999) 1715. 

\bibitem{osc1}
V. I. Yukalov, E. P. Yukalova, V. S. Bagnato, Phys. Rev. A
66 (2002) 043602

V. I. Yukalov, E. P. Yukalova, V. S. Bagnato, Phys. Rev. A
56 (1997) 4845



\bibitem{mit}S.  Inouye, M. R. Andrews, J. Stenger, H. J. Miesner,
D. M. Stamper-Kurn, W. Ketterle,   { Nature (London)} { 392}
(1998) 151;

P. Courteille, R. S. Freeland, D. J. Heinzen, F. A. van Abeelen,
B. J. Verhaar, 
 { Phys. Rev. Lett.}   
{ 81} (1998) 69.




 
 
\bibitem{3}D. S.  Jin,  J. R. Ensher, M. R.  Matthews, C. E.  Wieman,
E. A.
 Cornell,  {Phys. Rev. Lett.} { 77} (1996) 420;

M.-O.  Mewes, M. R.  Andrews, N. J.  van Druten, D. M. Kurn, D. S.
 Durfee, C. G.
 Townsend,  W.  Ketterle,    { Phys. Rev. Lett.} 
{ 77} (1996) 988.


 

 
\bibitem{8} F. Dalfovo, S. Giorgini, L. P. Pitaevskii,  S. Stringari,
Rev. Mod.  Phys. { 71} (1999) 463.






\bibitem{11} S. K.
  Adhikari, P.  Muruganandam,  {
J. Phys. B} { 35} (2002) 2831.


\bibitem{adx} S. K. Adhikari, Phys. Rev. A 66 (2002) 013611;

S. K. Adhikari, Phys. Rev. A 66 (2002) 043601.
 
 



\end{thebibliography}
 \end{document}